\documentclass{article}
\usepackage{hiph-preprint}
\usepackage{graphicx}
\volnumber{22} \issuenumber{1} \edyear{2005}                             
\frompage{000} \topage{000}                                              
\recrevdate{\today}                                                      
\def\rightmark{Properties of hadron and quark matter
studied with a molecular dynamics} 
\def\leftmark{Y. Akimura, T. Maruyama, N. Yoshinaga and S. Chiba}
 
\title{Properties of hadron and quark matter 
studied with a molecular dynamics} 
\authors{ 
{Yuka Akimura$^{1,2}$, Toshiki Maruyama$^{2}$, Naotaka Yoshinaga$^{1}$
and Satoshi Chiba$^{2}$ %
\index{1} 
\index{2} 
}\\[2.812mm]
{\normalsize
\hspace*{-8pt}$^1$ Department of Physics, Saitama University,\\ 
255 Shimo-Okubo, Sakura-Ku, Saitama-City, Saitama 338-8570, Japan\\[0.2ex] 
\hspace*{-8pt}$^2$ Advanced Science Research Center, Japan Atomic Energy
Research Institute,\footnote {Present name: Japan Atomic Energy Agency} \\ 
Tokai, Ibaraki 319-1195, Japan
}}
 
\abstract{
We study the hadron-quark phase transition in a molecular dynamics (MD)
of quark degrees of freedom.
The hadron state at low density and temperature,
and the deconfined quark state at high density and temperature
are observed in our model.
We investigate the equations of state and draw the phase-diagram
at wide baryon density and temperature range.
We also discuss the transport property,
e.g.\ viscosity, of $q\bar{q}$ matter.
It is found that the ratio of the shear viscosity to the entropy density
is less than one for quark matter.
}

 
\makeindex
\begin{document}
 
\maketitle
\setcounter{page}{1}

\section{Introduction}\label{intro}

Recently experimental results 
which suggest properties of quark matter 
at high temperature have been obtained \cite{1,2,3}.
They tell us that the shear viscosity of quark gluon plasma (QGP) 
is very small and the phase is liquid-like.
A theoretical study by quenched lattice QCD was performed on 
the matter viscosity at above the critical temperature \cite{4}.
The viscosity per entropy density $\eta/s$ was found to be much smaller
than 1, which indicates the matter is 
extremely smooth.
However, dynamics of quarks and anti-quarks, which seems very important, 
was not considered in the study.
In this paper we 
study the properties of QGP 
by employing a molecular dynamics (MD) which 
is a natural framework for describing 
both statical and dynamical aspects of the phase transition
\cite{5}.
We first show thermodynamical behavior of
quark matter at finite densities.
Then our preliminary results on the properties of 
matter at finite temperature and zero baryon density
are presented.

\section{Framework}\label{FW}  

We extend our previous MD model for zero temperature \cite{5}
to finite temperature.  
We prepare a cubic box having a length of 4 -- 10 fm, in which 
the same number of $u$ and $d$ quarks are contained,
and impose a periodic boundary condition.
To simulate quark many-body system at finite temperature,
we give appropriate initial values of 
position and momentum of $i$-th quark, ${\bf R}_i$ and ${\bf P}_i$,
and solve the following equations, 
\begin{equation}
\dot{{\bf R}_{\it i}}=\frac{\partial H}{\partial {\bf P}_{\it i}}
,\;\;\;
\dot{\bf P}_{\it i}=-\frac{\partial H}{\partial {\bf R}_{\it i}}
,\;\;\;
\end{equation}
where 
$H$ is a Hamiltonian used in Ref.\ \cite{5}.
The equations of motion are solved for several thousands fm/$c$
after the system reached equilibrium.

We simulate two kinds of quark matter, i.e.,
a matter with finite baryon density
and $q\bar{q}$ matter with zero baryon density.
The first one does not include $\bar{q}$ even at finite temperature
as a first trial.
In the calculation of the second case, $q\bar{q}$ matter, 
we slightly modify the Hamiltonian, i.e.\ meson-exchange 
terms \cite{5} are switched off.
The number of quarks and anti-quarks $N_q$ for $q\bar{q}$ matter
is determined as follows:
\begin{equation}
N_q=-\left(\frac{\partial \Omega}{\partial \mu}\right)\Bigg|_{\mu=0}, \ \
\Omega\equiv TV \int _0^{\infty}\frac{d^3k}{(2\pi)^3}\ln\left(
1+\exp\left[-\frac{\sqrt{k^2+m^2}-\mu}{T}\right]\right),  
\end{equation}
where $\mu$ is a chemical potential, $\Omega$ a thermodynamical potential,
$T$ a temperature, $V$ a volume and
$m$ denotes a quark mass of 300 MeV assumed in this calculation.
Note that this formula is for a free Fermi gas
and no effect of interaction is included.
Therefore the number of quarks and anti-quarks is not consistent
with the present MD simulation.


\section{Results and Discussion}\label{Res}

At first we discuss properties of quark matter.
Figure 1 shows temperature dependence of the energy
per baryon at $0.5\rho_0$ and $1\rho_0$.
The effective temperature is calculated from the momentum distribution \cite{6}
as,
\begin{equation}
T_{\rm eff}=
\frac{1}{N}\sum_{i=1}^{N}\frac{P_i}{2}\frac{\partial H}{\partial P_{\it i}}.
\end{equation}
In our MD simulation, quarks form baryons at low temperatures.
At a certain temperature 
%
%
a baryon-quark transition occurs and quarks are released from baryons.
This can be seen as the change of specific heat in Fig.\ 1.
We define the critical temperatures by maximum-specific-heat points
indicated by arrows.
The critical temperature becomes smaller as the density increases.
We draw a phase diagram of quark matter in Fig.\ 2.
Baryon phase is realized at the lower side of density and temperature,
and quark phase at the higher side.
Note that $q\bar{q}$ pair creation and annihilation are neglected in this model.
%
%
If thermally excited $q\bar q$ pairs were considered in the model,
the boundary of baryon and quark phases might move to the lower side
of density and temperature:
when the space is filled with more quarks and anti-quarks,
deconfinement of a hadron can occur more easily due to 
interaction with other quarks/anti-quarks in the environment.
%
\begin{figure}[t]
\begin{center}
\begin{minipage}{.42\textwidth}
\begin{center}
\includegraphics[width=0.8\linewidth]{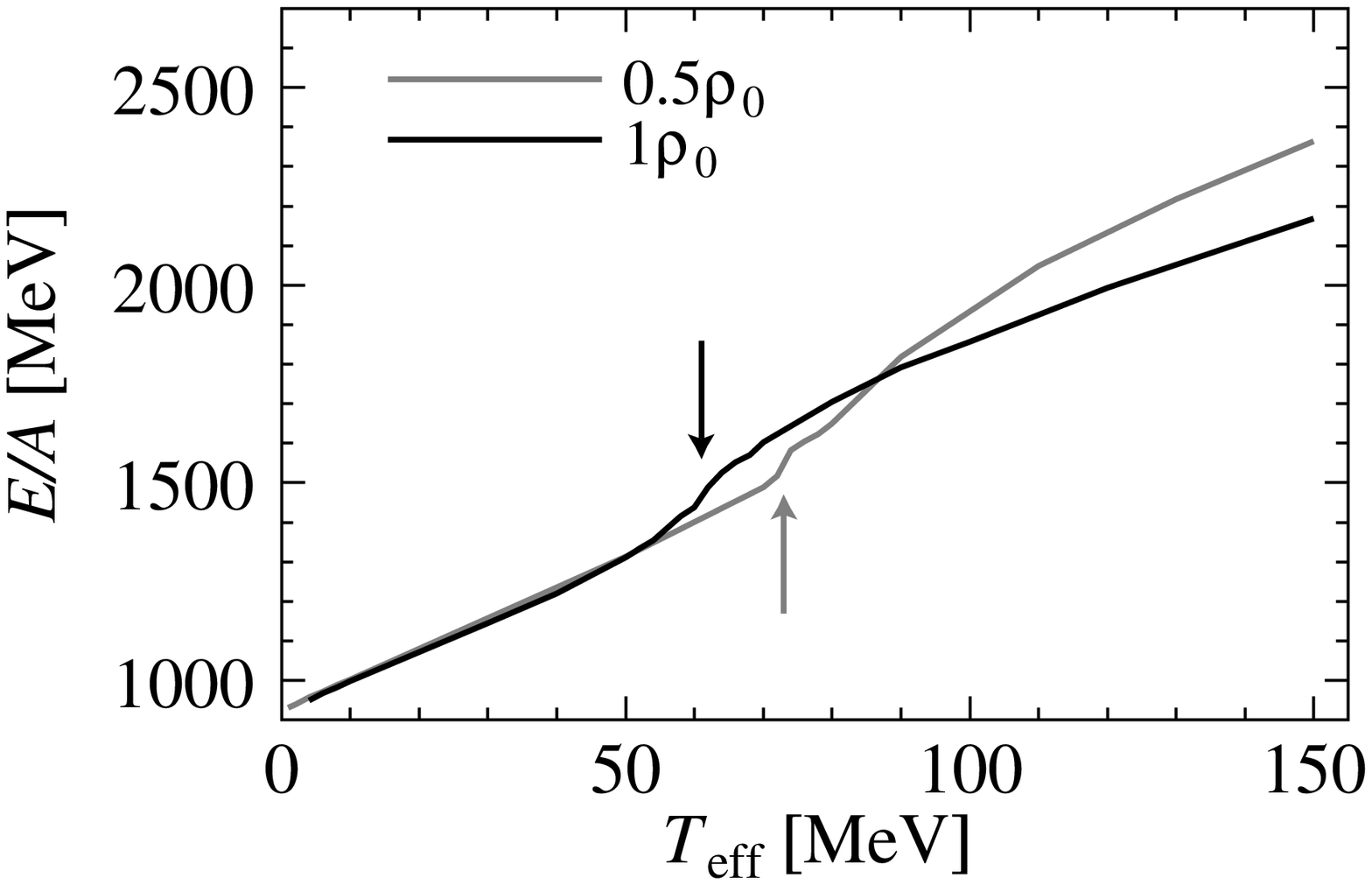}
\end{center}
\vspace*{-5mm}
\caption{
Temperature dependence of energy per baryon for 0.5$\rho_0$ (thin line)
 and 1$\rho_0$ (dark line).
The arrows indicate maximum-specific-heat points.}
\end{minipage}%
\hspace{5mm}
\begin{minipage}{.42\textwidth}
\begin{center}
\includegraphics[width=0.8\linewidth]{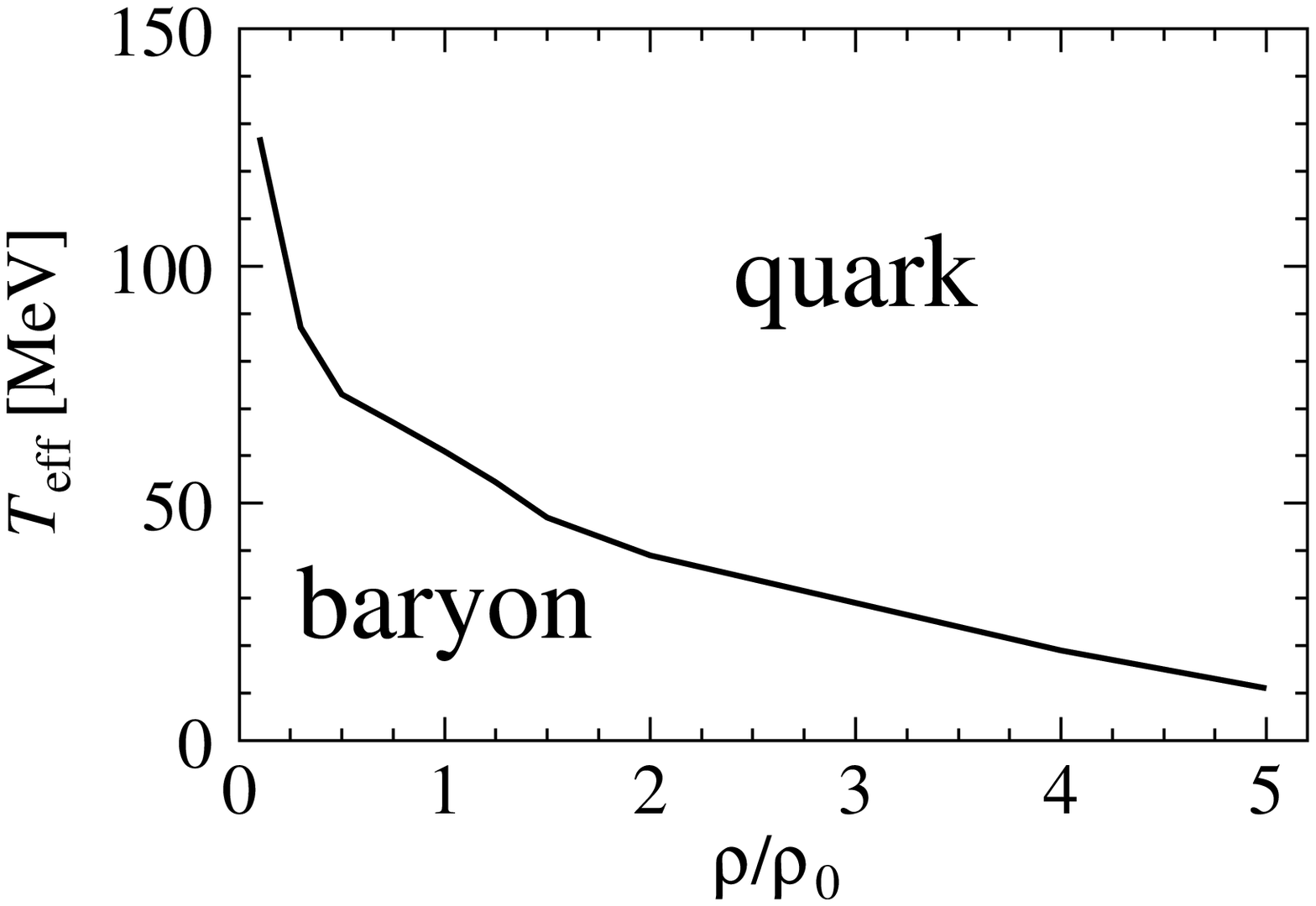}
\end{center}
\vspace*{-5mm}
\caption{
Phase diagram on the density-temperature plane.
The saturation density of the symmetric nuclear matter
is indicated by $\rho_0$.
}
\end{minipage}%
\end{center}
\end{figure}

Let us next discuss thermodynamical properties of $q\bar{q}$ matter.
The shear viscosity $\eta$ can be calculated
by using Green-Kubo relation
\begin{equation}
\eta=\frac{1}{T_{\rm eff}V}\int_{0}^{\infty} 
J_{xy}(t) \cdot J_{xy}(0)
dt,
\end{equation}
where 
$J_{xy}$ is the stress tensor in the $x$-$y$ direction \cite{8},
\begin{equation}
J_{xy}(t)=\sum_{i=1}^{N} x_i(t)\cdot \left(-\frac{\partial H}{\partial y_i}(t)\right)+
\frac{\partial H}{\partial P_{i,x}} (t) \cdot P_{i,y}(t).
\end{equation}
%
\begin{figure}[t]
\begin{center}
\begin{minipage}{.42\textwidth}
\begin{center}
\includegraphics[width=0.8\linewidth]{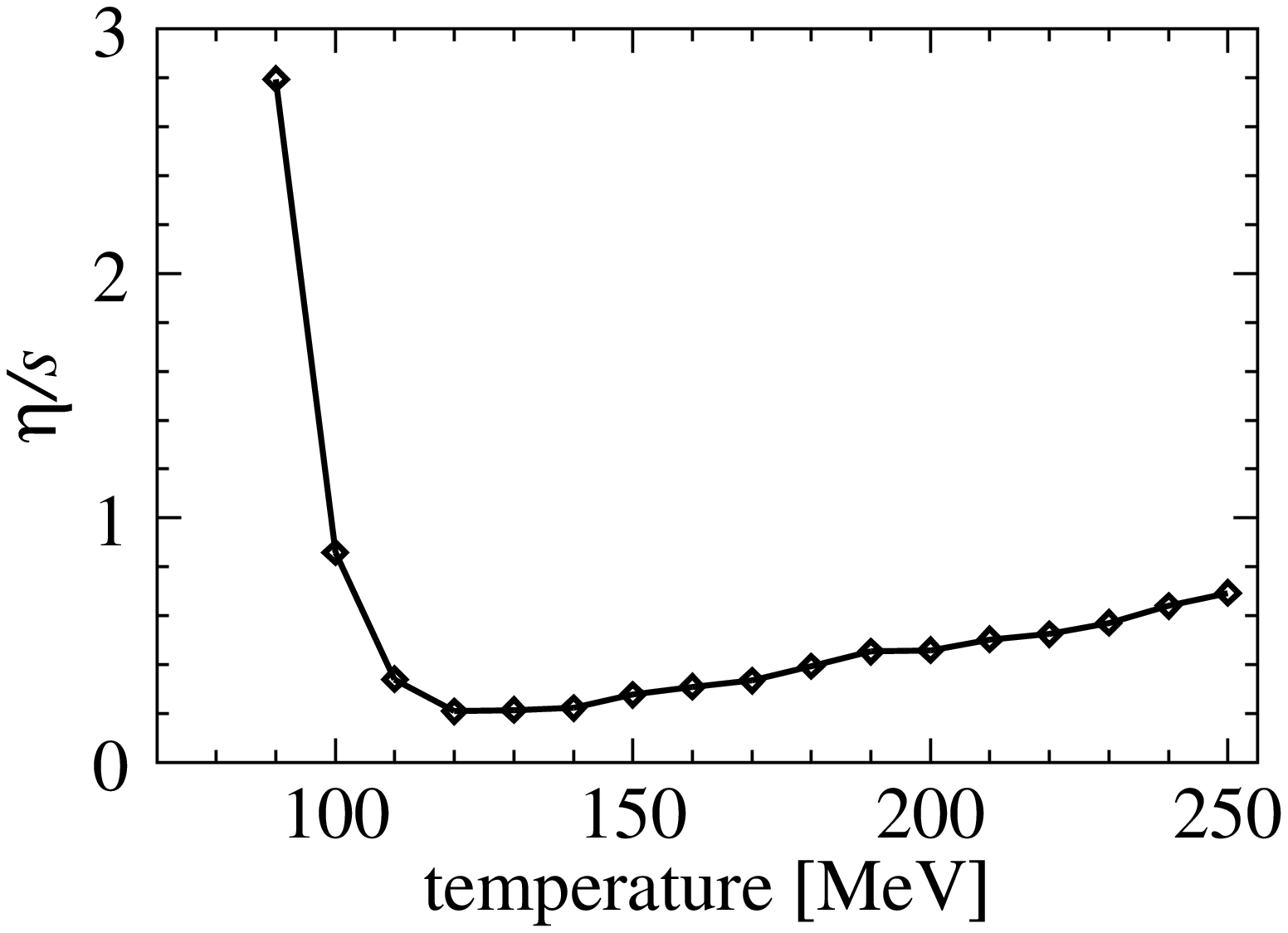}
\end{center}
\vspace*{-5mm}
\caption{Temperature dependence of the ratio of the shear viscosity
to the entropy density for $q\bar{q}$ matter.
}
\end{minipage}%
\hspace{5mm}
\begin{minipage}{.42\textwidth}
\begin{center}
\includegraphics[width=0.8\linewidth]{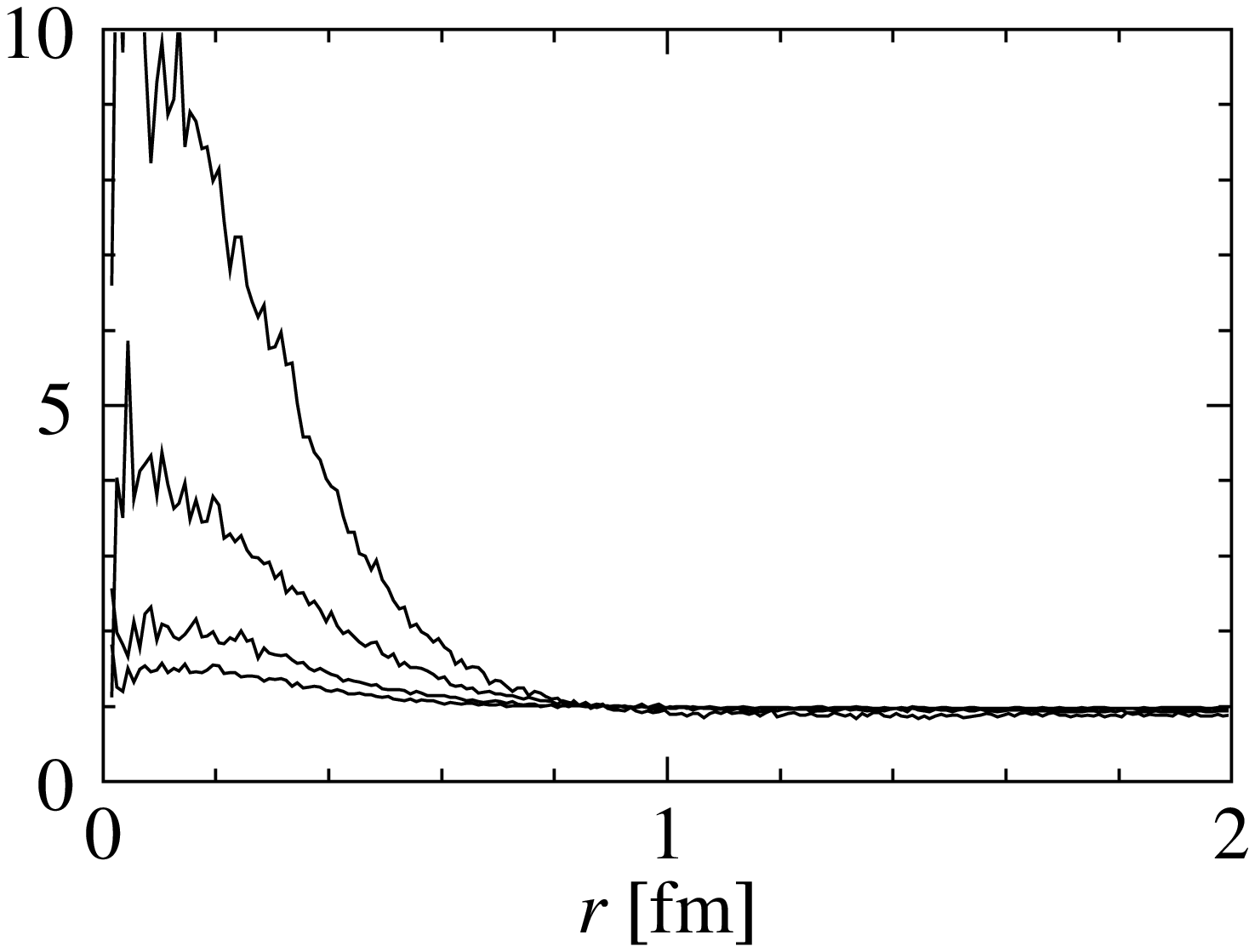}
\end{center}
\vspace*{-5mm}
\caption{Radial distribution function of $q\bar{q}$ matter
at 120, 140, 160 and 180 MeV from the top.
}
\end{minipage}%
\end{center}
\end{figure}
To discuss the ratio of the shear viscosity to the entropy density,
thermodynamic quantities in MD are calculated as 
\begin{equation}
s=\frac{\varepsilon+P}{T_{\rm eff}} , \ \ \
PV=NT_{\rm eff}+\frac{1}{3}
\sum_{i=1}^{N}{\bf R}_i \cdot 
\left(-\frac{\partial H}{\partial {\bf R}_i} \right) , 
\end{equation}
where $P$ is a pressure, $s$ is a entropy density and 
$\varepsilon$ is the energy density.
Simulation is performed by using a box size of 10 fm for low temperature
region and 4 fm for high temperature region.
Figure 3 shows the ratio of shear viscosity to the entropy density.
In the low temperature region,
where quarks form mesons,
the ratio $\eta/s$ is larger than one.
The value decreases as the temperature increases.
Around 130 MeV, where quarks are deconfined,
 $\eta/s$ takes a value smaller than one.
This value is of the same order as other studies \cite{4,9}.
Above 130 MeV, $\eta/s$ increases linearly.
This means that the motion of quarks are more dominant
than the interaction 
at high temperature.
If an effect of the asymptotic freedom were taken into account,
the ratio $\eta/s$ at high temperature would get larger.

Radial distribution functions of relative distance of two quarks
at 120, 140, 160 and 180 MeV
are shown in Fig.\ 4. 
They are normalized to one in the case of completely uniform distribution.
Bumps near zero distance are due to the meson-like correlation.
Change from the meson phase to the quark phase is 
seen as the temperature increases.
However, since the number of quarks and anti-quarks $N_q$ is not
consistently introduced in the present model,
the critical temperature or other quantities of phase transition
cannot be discussed quantitatively.

The results of this paper are still on a preliminary level and 
longer simulation time and larger box size are necessary 
for more quantitative discussions.
Investigation of viscosity for matter of finite baryon density 
is in progress.

Y. A.\ is grateful to H. Kaburaki and T. Hatsuda for their valuable discussions.

\vfill\eject
\end{document}